\documentclass{webofc}
\usepackage{graphicx}
\usepackage[varg]{txfonts}  
\begin{document}

\title{Minimizing plasma temperature for antimatter mixing experiments}

\author{\lastname{E.~D.~Hunter}\inst{1}\thanks{\email{eric.david.hunter@cern.ch}} \and
\lastname{C.~Amsler}\inst{1} \and
\lastname{H.~Breuker}\inst{2} \and
\lastname{S.~Chesnevskaya}\inst{1} \and
\lastname{G.~Costantini}\inst{3} \and
\lastname{R.~Ferragut}\inst{4,5} \and
\lastname{M.~Giammarchi}\inst{5} \and
\lastname{A.~Gligorova}\inst{1} \and
\lastname{G.~Gosta}\inst{3} \and
\lastname{H.~Higaki}\inst{6} \and
\lastname{Y.~Kanai}\inst{7} \and
\lastname{C.~Killian}\inst{1} \and
\lastname{V.~Kletzl}\inst{1} \and
\lastname{V.~Kraxberger}\inst{1} \and
\lastname{N.~Kuroda}\inst{8} \and
\lastname{A.~Lanz}\inst{1} \and
\lastname{M.~Leali}\inst{3} \and
\lastname{V.~M\"ackel}\inst{1} \and
\lastname{G.~Maero}\inst{5,9} \and
\lastname{C.~Mal\-bru\-not}\inst{10} \and
\lastname{V.~Mascagna}\inst{3} \and
\lastname{Y.~Matsuda}\inst{8} \and
\lastname{S.~Migliorati}\inst{3} \and
\lastname{D.~J.~Murtagh}\inst{1} \and
\lastname{Y.~Nagata}\inst{11} \and
\lastname{A.~Nanda}\inst{1} \and
\lastname{L.~Nowak}\inst{10} \and
\lastname{E.~Pasino}\inst{5,9} \and
\lastname{M.~Rom\'e}\inst{5,9} \and
\lastname{M.~C.~Simon}\inst{1} \and
\lastname{M.~Tajima}\inst{7} \and
\lastname{V.~Toso}\inst{4,5} \and
\lastname{S.~Ulmer}\inst{2} \and
\lastname{U.~Uggerh{\o}j}\inst{12} \and
\lastname{L.~Venturelli}\inst{3} \and
\lastname{A.~Weiser}\inst{1} \and
\lastname{E.~Widmann}\inst{1} \and
\lastname{T.~Wolz}\inst{10} \and
\lastname{Y.~Yamazaki}\inst{2} \and
\lastname{J.~Zmeskal}\inst{1}
\\
\lastname{(The ASACUSA-Cusp Collaboration)}}
\institute{Stefan Meyer Institute
\and
Ulmer Fundamental Symmetries Laboratory, RIKEN
\and
Dipartimento di Ingegneria dell'In\-formazione, Universit\`a degli Studi di Brescia and INFN Pavia
\and
Politechnico di Milano
\and
INFN Milano
\and
Graduate School of Advanced Science and Engineering, Hiroshima University
\and
Nishina Center for Accelerator-Based Science, RIKEN
\and
Institute of Physics, the University of Tokyo
\and
Dipartimento di Fisica, Universit\`a degli Studi di Milano
\and
Experimental Physics Department, CERN
\and
Department of Physics, Tokyo University of Science
\and
Department of Physics and Astronomy, Aarhus University
}
%\author{\lastname{(The ASACUSA Collaboration)}}

\abstract{The ASACUSA collaboration produces a beam of antihydrogen atoms by mixing pure positron and antiproton plasmas in a strong magnetic field with
a double cusp geometry. The positrons cool via cyclotron radiation inside the cryogenic trap. Low positron temperature is essential for increasing the fraction of antihydrogen atoms which reach the ground state prior to exiting the trap. Many experimental groups observe that such plasmas reach equilibrium at a temperature well above the temperature of the surrounding electrodes. This problem is typically attributed to electronic noise and plasma expansion, which heat the plasma. The present work reports anomalous heating far beyond what can be attributed to those two sources. The heating seems to be a result of the axially open trap geometry, which couples the plasma to the external (300 K) environment via microwave radiation.}

\maketitle

\section{Introduction}
\label{sec:intro}
\begin{figure}[h]
\centering
\includegraphics[width=\linewidth,clip]{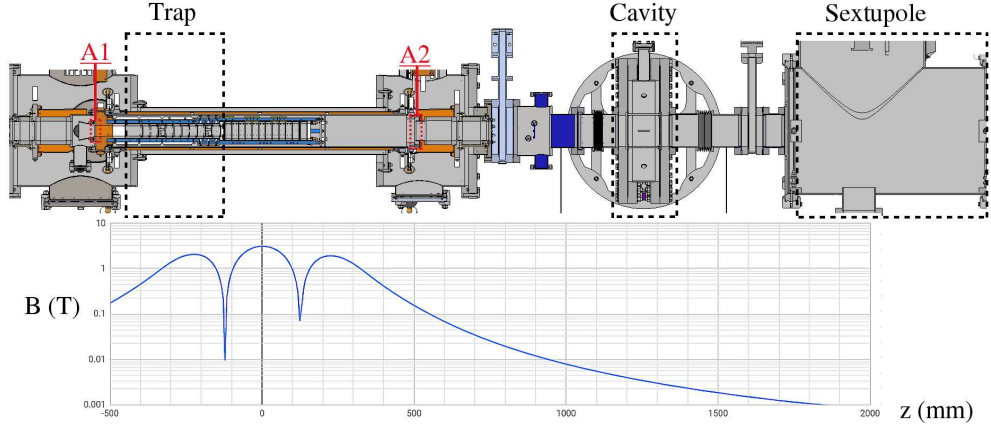}
\caption{The Cusp experiment. Positrons and antiprotons are loaded into ``Trap'' from other traps further upstream (left, not shown). The antimatter plasmas are manipulated, cooled, and mixed together to form antihydrogen. The two cusps in the magnetic field $B$ help to focus low energy, low-field-seeking antiatoms into a spin-polarized beam traveling to the right at approximately $1000\,\mathrm{m/s}$. The frequency of the microwaves applied to ``Cavity'' is varied, while monitoring the number of antiatoms that are refocused by ``Sextupole'' onto a scintillator target (right, not shown). When the microwave frequency is close to the hyperfine splitting ($\nu_{HFS}\approx 1.420\,\mathrm{GHz}$), low-field-seeking atoms passing through ``Cavity'' become high-field-seekers. High-field-seekers are defocused by ``Sextupole''. Thus, the transition frequency is detected as a dip in the signal on the scintillator. ``A1'' and ``A2'' are apertures, necessary for admitting particles and letting antiatoms escape from the trap. These apertures also admit room-temperature microwave radiation, which may heat the plasma.}
\label{fig:cusp}
\end{figure}

The ASACUSA-Cusp project aims to measure the ground state hyperfine splitting of antihydrogen in a magnetic field-free region with a precision of $1\,\mathrm{ppm}$ \cite{malb:18}. The hyperfine measurement is favored among possible probes of CPT violation because of the way it depends on the spin of the positron-antiproton system \cite{bluh:99}. The experimental concept is sketched in Fig.~\ref{fig:cusp}.

The number of antiatoms produced in mixing attempts so far is insufficient to perform a precision measurement; the ground state production rate must increase by about a factor of one hundred \cite{kolb:21}. For increasing the antihydrogen yield, the most important parameter is positron temperature. A $20\,\mathrm{K}$ plasma should yield x100 more ground state antiatoms than the equivalent $200\,\mathrm{K}$ plasma \cite{radi:14}. In other words, the principle obstacle to this measurement appears to be the unfortunate fact that plasma in the Cusp trap tends to equilibrate closer to $200\,\mathrm{K}$ than to $20\,\mathrm{K}$. 

An electron or a positron plasma in a strong magnetic field will cool via cyclotron radiation at a rate $\Gamma = 0.26\,\mathrm{s^{-1}}\times B[\mathrm{T}]^2$ according to
\begin{equation}
    \frac{dT}{dt}=-\Gamma(T-T_b)+H
    \label{eq:newton}
\end{equation}
where $T$ is the plasma temperature, $t$ is the time, $T_b$ is the temperature of the thermal bath, and $H$ is the heating rate due to plasma expansion and electrode noise. If $\Gamma$, $H$, and $T_b$ are constant, $T$ should asymptote exponentially to the value 
\begin{equation}
    T_f=T_b+H/\Gamma
    \label{eq:Tf}
\end{equation}
If $T_b$ is taken as the trap temperature $T_t = 35\,\mathrm{K}$, then a plasma temperature $T_f\sim 60 \,\mathrm{K}$ is a realistic expectation for typical values of $H$. 

The factors contributing to $H$ were minimized in the present trap by using low-noise electronics, low-pass filtering, and corrective trap alignment. It was therefore surprising to find that plasma with seemingly low $H$ (see Section~\ref{sec:heating}) relaxes to a stubbornly high temperature $T_f\approx 170\,\mathrm{K}$ (see Section~\ref{sec:shield}). Much of the 2021 experimental campaign at CERN was dedicated to understanding and addressing this discrepancy. 

Section~\ref{sec:shield} describes experiments using a movable thermal shield to block room-temperature radiation entering the cryogenic trap at the downstream end, equivalent to ``A2'' in Fig.~\ref{fig:cusp}. The base plasma temperature was $20\,\mathrm{K}$ lower when the shield was closed. This result motivated the installation of a similar shield to block the upstream end, equivalent to ``A1''. The mount for this shield constituted a 75\% reduction in the area of ``A1'' and a coincident $40\,\mathrm{K}$ reduction in minimum plasma temperature was observed. The change in $T$ was entirely independent of whether the small upstream shield was open or closed. 

Sections~\ref{sec:heating} and~\ref{sec:temperature} describe a study of expansion heating, cyclotron cooling, and final temperature as a function of the plasma's location in the trap. The study permitted a quantitative evaluation of $T_b \approx T_t + 70\,\mathrm{K}$ in the configuration with both shields installed.

In light of the accumulating evidence that the thermal bath coupling to the plasma had a room-temperature component, copper meshes were installed to completely block microwave radiation through ``A1'' and ``A2''. The resulting reduction in base plasma temperature was immediate and significant. However, those results were obtained after the EXA conference and will be reported elsewhere.

\section{Movable radiation shield}
\label{sec:shield}

A cryogenic, movable thermal shield was first used for plasma temperature studies in the ALPHA-1 experiment \cite{jenk:08}. The magnetically actuated shield could close the entrance to the trap, preventing microwaves (and possibly residual gas) from entering. Initial observations indicated that the plasma temperature was lower when the shield was closed (J. Fajans, private communication). Povilus \cite{povi:15} and Carruth \cite{carr:18} installed similar devices in other experiments, however, without reproducing the effect on plasma temperature observed in the previous apparatus. The shield used in the present work had been used in earlier versions of the Cusp trap for achieving lower electrode temperature \cite{enom:11}, and was removed c. 2016. It was re-installed in 2021 to see if the plasma temperature effect observed elsewhere could be reproduced in a more recent version of the Cusp trap.

\begin{figure}[h]
  \begin{minipage}{0.12\textwidth}
        \centering
        \includegraphics[width=\linewidth]{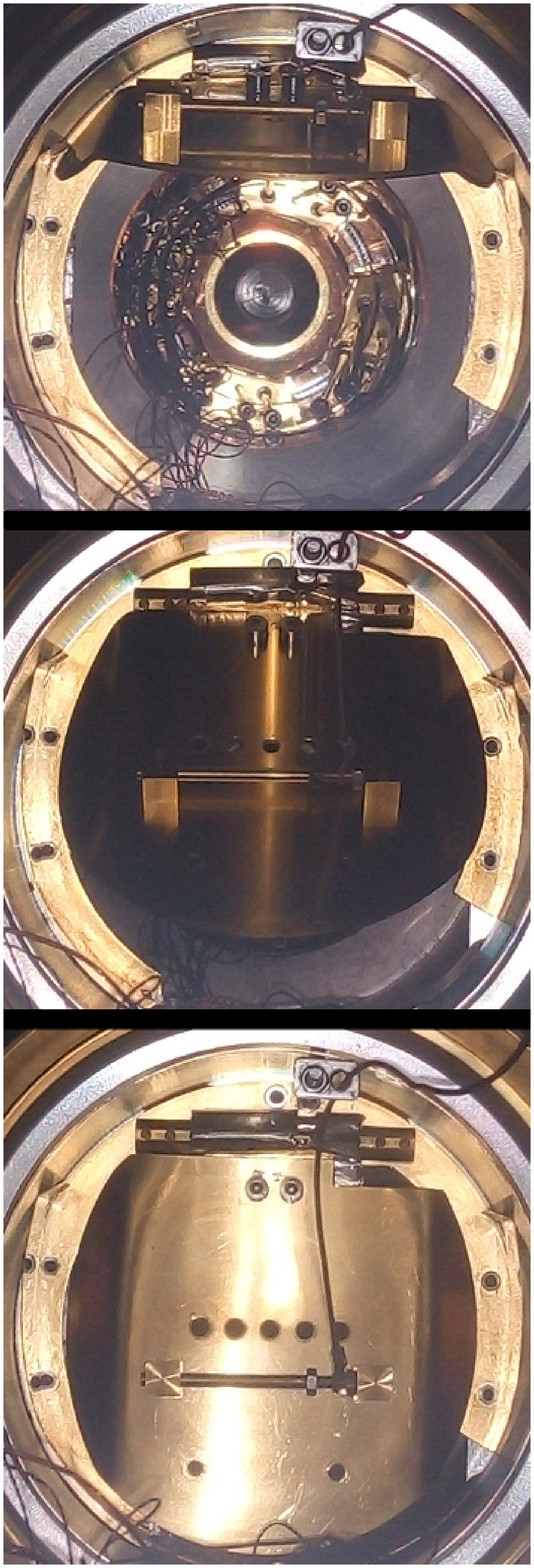}
    \end{minipage}\hfill
    \begin{minipage}{0.38\textwidth}
        \centering 
        \includegraphics[width=\linewidth]{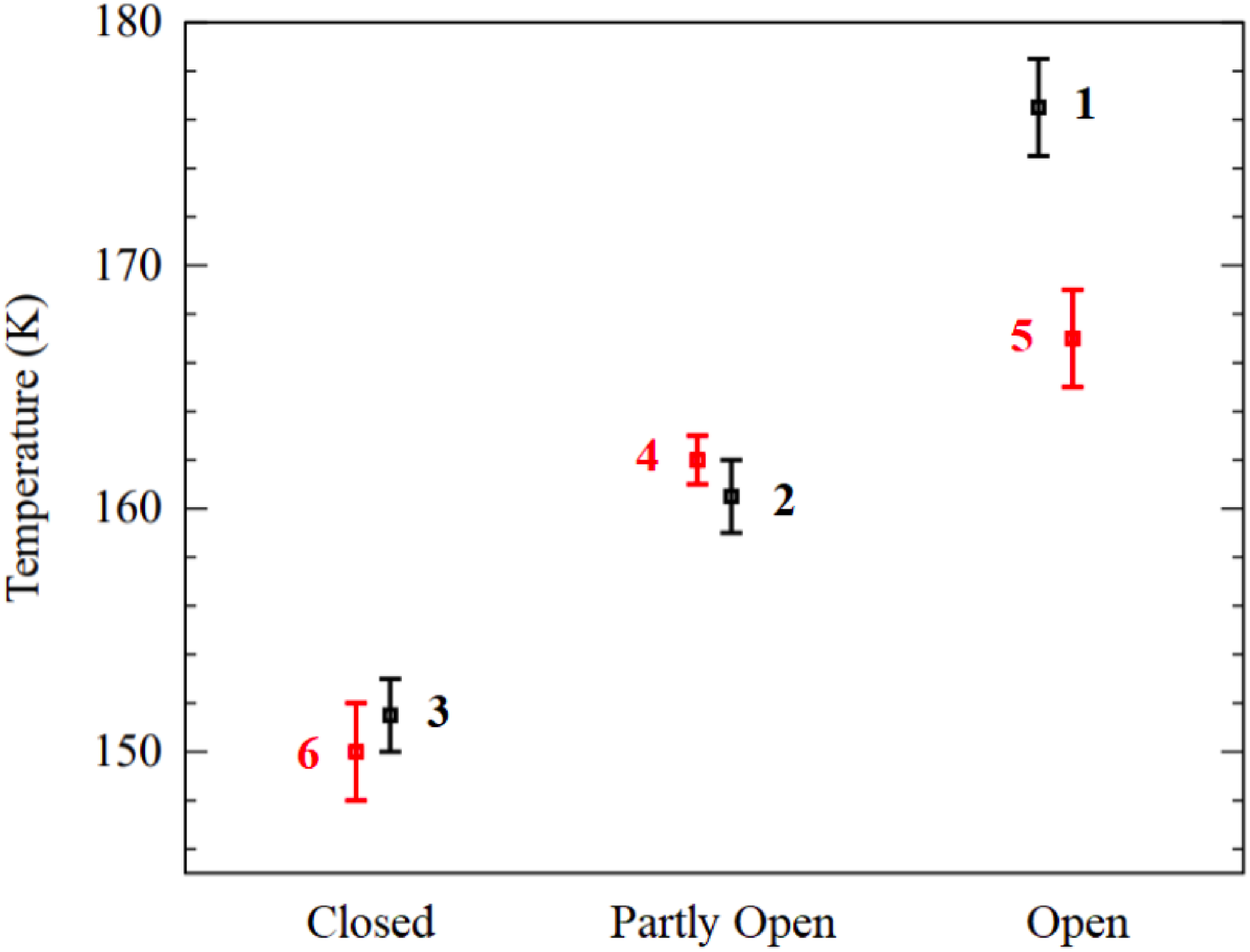}
    \end{minipage}\hfill
    \begin{minipage}{0.45\textwidth}
        \centering 
        \includegraphics[width=\linewidth]{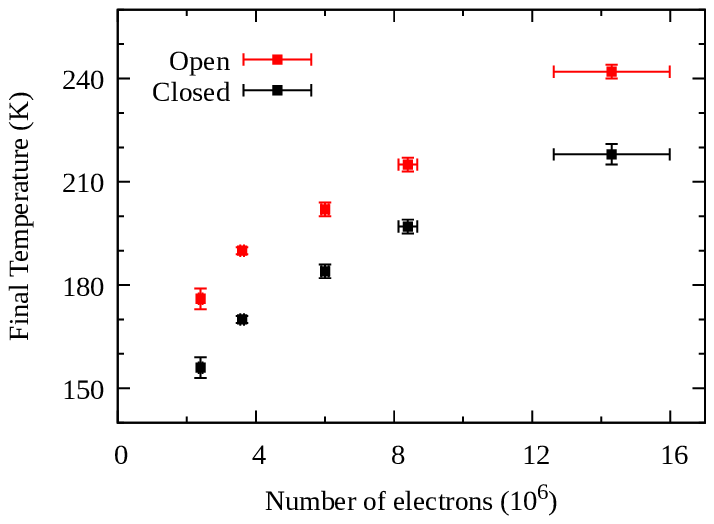}
    \end{minipage}
\caption{Left: Photos of the downstream shield taken through a vacuum window. The shield is open in the top photo, partly open in the middle photo, and closed in the bottom photo. Notice that the shield is still mostly closed in the ``partly open'' configuration; the trap (visible as a small circle in the top photo, having diameter roughly 1/6 the size of the photo) is still completely covered by the shield. Middle: Plasma temperature after cooling with the shield in the three positions shown at the left. Each point gives the mean and standard deviation of $T$ for 15 different plasmas. The numbers next to the points indicate the order in which the data was taken; to further emphasize the ordering, the first three points in the series are black and the last three are red. Right: Plasma temperature as a function of the number of electrons in the plasma. The temperature is $20\,\mathrm{K}$ higher when the downstream shield is open, independent of the number of electrons in the plasma.}
\label{fig:higaki}
\end{figure}

Figure~\ref{fig:cusp} shows the current version of the Cusp trap (winter 2021). The experiments reported here, however, were performed in an older version (summer 2021), with nearly identical electrodes but in a slightly different vacuum chamber. The downstream end of this chamber was protected from thermal radiation by the heavy copper shield mentioned above. The shield was located roughly $150\,\mathrm{mm}$ downstream of ``A2''. The shield could be opened by pulling on a braided steel cable, which is visible at the top of the photos in Fig.~\ref{fig:higaki}. The shield was not cryogenic because the downstream coldhead was not working at the time of these measurements. While the trap electrodes were at $35\,\mathrm{K}$, the axial center of the cold bore was at $60\,\mathrm{K}$ and the downstream shield was at $200\,\mathrm{K}$. These temperatures did not change by more than a degree over the course of weeks of experimentation, nor were they changed by opening or closing the downstream shield, whether for minutes or hours.

A plasma containing $N\approx 3\times 10^6$ electrons was cooled under a long electrode at the magnetic field maximum. The final temperature of the plasma was found to depend on the state of the downstream thermal shield according to the middle panel of Fig.~\ref{fig:higaki}. A position of the shield was found for which the temperature increment associated with the shield position was roughly half that found with the shield fully open. This state, labeled ``Partly Open'', was photographed from several angles to judge whether it permitted a line-of-sight path into the trap. No such path existed. It is hard to imagine what could get past the ``Partly Open'' shield and heat the plasma other than microwaves.

The right side of Fig.~\ref{fig:higaki} shows that the temperature increment with the shield open was the same for plasmas with significantly different $N$ (and different $T_f$). This provides further evidence that the ``anomalous heating'' has nothing to do with the plasma and may be equated to a simple offset in $T_b$ in Eq.~\ref{eq:Tf}. 

In response to these observations, another shield was installed at the upstream end of the trap. The temperature of the upstream shield mount was $65\,\mathrm{K}$. The temperature of the electrodes remained $35\,\mathrm{K}$. The shield mount reduced the upstream aperture at ``A1'' from $40$ to $20\,\mathrm{mm}$. When nearly the same sequence of plasma manipulations for generating the data in Fig.~\ref{fig:higaki} was run again in this configuration, the final plasma temperature was found to be approximately $40\,\mathrm{K}$ lower (see Section~\ref{sec:temperature}). However, closing the upstream shield did not have any measurable effect on the temperature, despite much effort to ensure that the shield opened in less than $100\,\mathrm{ms}$ so that the plasma would not warm up while the shield was opening for the plasma temperature diagnostic (see Section~\ref{sec:heating}). The remainder of this work is concerned with data obtained in this configuration, with the downstream shield closed and the upstream shield open but the aperture reduced by the shield mount.

\section{Conventional plasma heating}
\label{sec:heating}

The charges in the plasma can gain kinetic energy (plasma heating) from two principle sources. First, when the plasma expands, potential energy is directly converted into kinetic energy as charges push away from each other. Second, plasma modes can be stimulated by radiofrequency noise on the electrodes, and these modes can be damped kinetically, transferring the energy from a single plasma mode into the thermal ensemble. 

\begin{figure}[h]
  \begin{minipage}{0.48\textwidth}
        \centering
        \includegraphics[width=\linewidth]{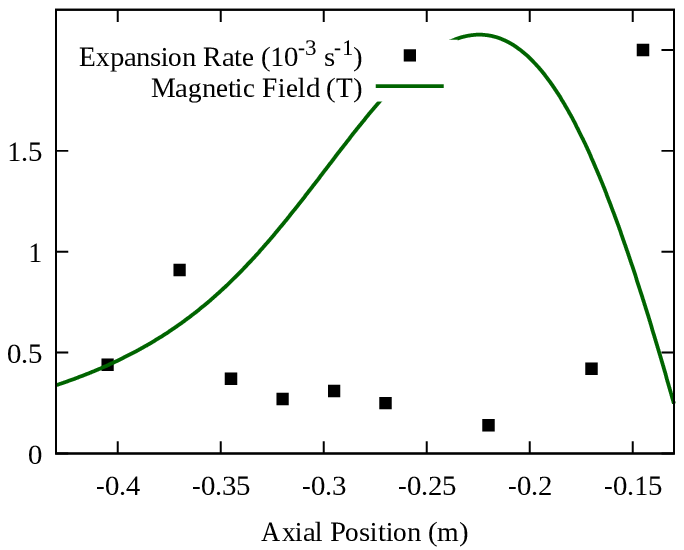}
    \end{minipage}\hfill
    \begin{minipage}{0.48\textwidth}
        \centering 
        \includegraphics[width=\linewidth]{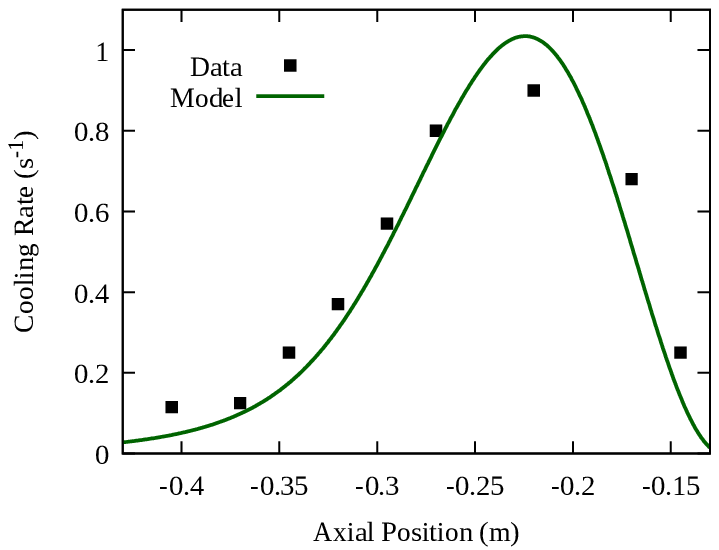}
    \end{minipage}
\caption{Left: Plasma expansion rate and axial magnetic field. Plasma containing $N\approx 3\times 10^6$ electrons is held for $1<t<300\,\mathrm{s}$. The expansion rate is the reciprocal of the amount of time required for the plasma radius to double. Plasma radius is determined from the standard imaging diagnostic \cite{peur:93}. Right: Cooling rate compared to the formula for $\Gamma$ given in Eq.~\ref{eq:newton}. Cooling rate is taken as the best exponential fit to temperature vs. time data, as the plasma cools from $T\sim 6000\,\mathrm{K}$ to $T\sim200\,\mathrm{K}$. Plasma temperature is determined from the standard temperature diagnostic \cite{beck:90}.}
\label{fig:expansion}
\end{figure}

Plasma expansion heating is reduced by reducing the plasma density, reducing the plasma radius, and reducing the expansion rate of the plasma. In a well-built trap with good vacuum, plasma expansion is primarily due to asymmetry transport, most often in the form of a misalignment between the magnetic field's axis of symmetry and the trap electrodes' axis of symmetry. That alignment can be roughly diagnosed by holding plasma in different axial locations ($z$ coordinate) of the trap and slowly dumping it onto an imaging detector. If the plasma is cold and dense, most of it will come out of the radial center of the trap (at that $z$ position), then follow the magnetic field lines to the detector. If a given magnetic field line crosses the trap center in only one of the two locations tested, then the imaged plasmas will appear in different positions on the detector. By rotating the electrode stack so as to minimize the difference between the two detected positions, the expansion rate was reduced to $1/\tau < 0.001\,\mathrm{s}^{-1}$ for most of the trap (see Fig.~\ref{fig:expansion}). This corresponds to a heating rate $H<7\,\mathrm{K/s} \times n r_p^2$, where the plasma density $n$ and radius $r_p$ are referred to $1\,\mathrm{mm}$ and $1\times 10^8\,\mathrm{cm}^{ -3}$, respectively. (This heating rate is calculated as $(dU/dt) / (3/2)Nk_B$, where $dU/dt=\omega_r dP/dt$, $N$ is the number of particles, $k_B$ is Boltzmann's constant, $\omega_r$ is the plasma rotation rate, and $P$ is the total canonical angular momentum of the plasma. See Eqs. (37) and (64) of Ref.~\cite{dani:15}.)

Heating from electrode noise occurs when one of the normal modes of the plasma is resonant with a Fourier component of the voltage applied to the confining electrodes. Noise near the resonant frequencies of the lowest order axial modes, which are typically in the frequency range $1<f<50\,\mathrm{MHz}$, tends to heat the plasma the most, probably because these are the highest-$Q$ modes and are the easiest to stimulate in the azimuthally symmetric trap geometry. These frequencies are excluded from the trap in three stages: (1) the electrode amplifiers have a bandwidth of roughly $1\,\mathrm{kHz}$, (2) the filterbox mounted directly onto the air-vacuum feedthru contains single-pole lowpass filters with time constant $RC=10\,\mathrm{\mu s}$, (3) cryogenic two-pole lowpass filters with time constant $RC=8\,\mathrm{\mu s}$ are mounted inside the trap, $1\,\mathrm{cm}$ away from the electrodes. The rotating-wall electrodes (middle of the trap) and the catching electrode (upstream end of the trap) have an AC bypass to allow high frequency signals to enter. The bypass consists of an AC-coupled line on the feedthru filterboard, which is normally grounded via a relay. The signals traverse the vacuum chamber on Lakeshore Quad-Twist line, with usually two of the four twisted wires being grounded at both ends to reduce magnetic pickup. Altogether, the noise rejection seems to be very effective. A $5\,\mathrm{V}$ broadband noise source ($1<f<20\,\mathrm{MHz}$) was applied to various instruments and feedthrus entering the trap, including the MCP, the electron source, and the field ionizers (of which the wiring runs parallel to the electrode wiring internally). The only place where the noise source produced a measurable change in plasma temperature was the AC input to a rotating wall (AC bypassed) electrode, and then, only when the relays were closed. These relays are normally open except when the rotating wall is being applied. 

For the trap described here, then, heating due to electrode noise seems to be negligible. Heating due to plasma expansion was calculated above to be roughly $10\,\mathrm{K/s}$ for most plasmas. Taken together with the cooling rate $\Gamma \gtrsim 0.4\,\mathrm{s}^{-1}$ and the trap temperature $T_t=35\,\mathrm{K}$, one would expect plasma to cool to $T<60\,\mathrm{K}$ in the worst case, and $T<50\,\mathrm{K}$ at the maximum magnetic field. That is not what was observed.

\section{Plasma temperature}
\label{sec:temperature}

Equation~\ref{eq:Tf} can be used to estimate $H$ and $T_b$ by observing the cooling rate $\Gamma$ and final temperature $T_f$ for plasma cooled in different locations. One assumes a value for either $H$ or $T_b$ and solves Eq.~\ref{eq:Tf} for the other. The results are given in Table~\ref{tab:positions}.

The plasma used to obtain the results in Table~\ref{tab:positions} had the following parameters: $N=3.5\times10^6$ electrons, $r_p=0.7\,\mathrm{mm}$, $n=3\times10^8\,\mathrm{cm}^{-3}$, $L_p\approx 4\,\mathrm{cm}$. Because the plasma length and position are different from those used to obtain the data in Fig.~\ref{fig:expansion}, the expansion rates were scaled by the factor $(L_p/B)^2$ \cite{dris:83} and averaged accordingly before applying the formula for the heating rate given in the previous section.

%http://vps2007266.fastwebserver.de/PLASMA/319
\begin{table}[h]
\centering
\caption{For each $z$ position, a complete $T$ vs. $t$ curve was taken so as to obtain experimental values for $\Gamma$ and $T_f$. $H_t$ is the heating rate obtained from Eq.~\ref{eq:Tf} by assuming that $T_b=T_t=35\,\mathrm{K}$. $H_e$ is the heating rate calculated from the plasma expansion rates given in Fig.~\ref{fig:expansion}. $T_{be}$ is the value of $T_b$ obtained assuming heating is only due to expansion, i.e. setting $H=H_e$ in Eq.~\ref{eq:Tf}.}
\label{tab:positions}
\begin{tabular}{llllll}
$z\ (\mathrm{m})$    & $\Gamma\ (\mathrm{s}^{-1})$  & $T_f\ (\mathrm{K})$ & $H_t\ (\mathrm{K/s})$   & $H_e\ (\mathrm{K/s})$    & $T_{be}\ (\mathrm{K})$  \\\hline
-0.38 & 0.09    & 300   & 24    & 29    & -20   \\
-0.32 & 0.35    & 130   & 33    & 13    & 92    \\
-0.27 & 0.85    & 105   & 60    & 12    & 91    \\
-0.22 & 0.90    & 118   & 75    & 6     & 112   \\
-0.17 & 0.35    & 130   & 33    & 49    & -11   \\
\end{tabular}
\end{table}

The data in Table~\ref{tab:positions} leads to three conclusions.
\begin{itemize}
    \item[1] The minimum plasma temperature is about $70\,\mathrm{K}$ higher than the expectation derived in the previous section, where it was assumed that $T_b=T_t$.
    \item[2] The higher temperatures at the two $z$ extremes can be predicted by $T_f=H_e/\Gamma$; $T_{be}$ is not well defined for these two cases because the relative contribution of $H_e/\Gamma$ is much larger.
    \item[3] In the middle of the trap, where the cooling rate is high and the expansion rate is low, the expansion heating is entirely insufficient to account for the observed plasma temperature unless $T_b$ is greater than $T_t$. The best estimate for $T_{be}$ is about $100\,\mathrm{K}$ in all three cases, which is about $70\,\mathrm{K}$ higher than $T_t$. This is consistent with conclusion (1). 
\end{itemize}

These observations, along with those reported in Section~\ref{sec:shield}, are simply explained by the hypothesis that the temperature of the thermal bath ($T_b$) is generally higher than the temperature of the trap ($T_t$). For the purposes of cyclotron cooling, the thermal bath includes every electromagnetic mode which couples to the cyclotron motion of the plasma particles. Many of these modes are propagating, or else have a non-vanishing component outside the cryogenic region. In either case, mode energy is damped by currents on the walls of the vacuum chamber outside the cryogenic region. This raises the effective temperature of the radiation environment seen by the plasma and limits the minimum achievable plasma temperature for any trap with open ends.

\section*{Acknowledgments}
This work was supported by the European Research Council under European Union’s Seventh Framework Programme (FP7/2007-2013) / ERC Grant Agreement (291242); the Austrian Ministry of Science and Research, Austrian Science Fund (FWF) W1252-N27 and P 34438; the JSPS KAKENHI Fostering Joint International Research B 19KK0075; the Grant-in-Aid for Scientific Research B 20H01930;  Special Research Projects for Basic Science of RIKEN; Università di Brescia and Istituto Nazionale di Fisica Nucleare; and the European Union’s Horizon 2020 research and innovation programme under the Marie Skłodowska-Curie grant agreement No 721559.

\bibliography{bib}

\end{document}